\documentclass[aps,prb,twocolumn,superscriptaddress,showpacs]{revtex4}

\usepackage{graphicx}

\begin{document}

\title{Effect of inhomogeneous magnetic flux on double-dot Aharonov-Bohm Interferometer}

\author{Zhi-Ming Bai}
\email{baizhiming@tsinghua.org.cn}
\affiliation{Physics Department,
Hebei Science and Technology University, Shijiazhuang 050018,
China} \affiliation{Department of Physics, Tunghai University,
Taichung, Taiwan}

\author{Min-Fong Yang}
\email{mfyang@mail.thu.edu.tw}
\affiliation{Department of Physics,
Tunghai University, Taichung, Taiwan}

\author{Yung-Chung Chen}
\email{ycchen@mail.thu.edu.tw}
\affiliation{Department of Physics,
Tunghai University, Taichung, Taiwan}

\date{\today}

\begin{abstract}
The influence of the inhomogeneous distribution of the magnetic
flux on quantum transport through coupled double quantum dots
embedded in an Aharonov-Bohm interferometer are investigated. We
show that the effective tunnelling coupling between two dots can
be tuned by the magnetic flux imbalance threading two AB subrings.
Thus the conductance and the local densities of states become
periodic functions of the magnetic flux imbalance. Therefore,
transport signals can be manipulated by adjusting the magnetic
flux imbalance. Thus accurate control of the distribution of the
magnetic flux is necessary for any practical application of such
an Aharonov-Bohm interferometer.
\end{abstract}

%
%

\maketitle

\section{Introduction}

Due to recent advances in nanotechnologies, quantum transport
through ultra-small quantum dots (QD) has drawn considerable
interests in the last decades.\cite{books} In such small
structures with geometrical dimensions smaller than the elastic
mean free paths, electron transport is ballistic and its phase
coherence can be sustained. To probe the coherence, interference
experiments, most notably Aharonov-Bohm (AB) interferometry, are
needed. The presence of conductance oscillations as a function of
magnetic flux has been experimentally demonstrated for AB
interferometers with one
QD.\cite{Yacobi95,Schuster97,Ji00,Wiel00,Kobayashi,Kobayashi-fano}
In Ref.~\onlinecite{Kobayashi-fano}, mesoscopic Fano effect with
complex Fano's asymmetric parameters is observed, and it is shown
that Fano effect can be a powerful tool to investigate the
electron phase variation in such mesoscopic transport.

Recently, the AB interferometer containing two {\it coupled} QD's
with a QD inserted in each arm has also been studied
experimentally.\cite{holleitner01,holleitner02,blick03,Sigrist03}
While there have already been many works for the AB interferometer
containing two QD's, most of them consider only the system {\it
without} direct coupling between dots. Particular interest in the
coupled system lies in its potential application in quantum
communication,\cite{Loss00} because entanglement of electrons is
possible in the presence of direct tunnelling between dots. Just
like the molecule of two atoms, two coupled QD's can form the
bonding and antibonding states. Therefore, such an AB
interferometer can also be used to probe the phase coherence of
the bonding between dots. Moreover, the possibility to control
each of two QD's separately increases the dimension of the
parameter space for the transport properties as compared to their
single-dot AB counterparts. Thus it can be considered as the
starting point of the study for experimentally unexplored region.
Motivated by these experimental works, theoretical investigations
for such a system have just begun.\cite{Jiang02,Kang02} It is
noted in Ref.~\onlinecite{Jiang02} that the interdot tunnelling
divides the AB interferometer into two coupled subrings, and the
total magnetic flux through the device is composed of magnetic
flux through two subrings. If the applied magnetic field is
non-uniform and/or the construction of the AB interferometer is
asymmetric, the magnetic flux threading two subrings can in
general be different.\cite{note-1} The possibility of non-uniform
distribution of magnetic flux is not taken into account in
Ref.~\onlinecite{Kang02}.

In this paper, the AB interferometer containing two {\it coupled}
QD's is investigated, and we focus our attention on the effect of
inhomogeneous magnetic flux. While this effect has been studied in
Ref.~\onlinecite{Jiang02} by solving numerically the modified rate
equations, they consider only a special case with integer values
of magnetic flux ratio of two subrings. Our aim is to provide
general analytic expressions of the conductance and the local
densities of states, which may serve as guides for the ongoing and
future experimental endeavor.\cite{note-2}

By solving exactly a simple model system, we derive general
formulas of the conductance and the local densities of states,
which includes most of the previous
results.\cite{Kang02,Jiang02,Guevara} Moreover, we find that
non-uniform distribution of the magnetic field piercing the AB
interferometer will give an important influence on electrons
transport. Thus complex characteristic transport features can
occur, which can easily be manipulated by applied gate voltages
and magnetic flux. First, the magnetic flux imbalance contributes
a phase factor on the tunnelling coupling. Thus the overlapping of
the dot's wave functions can be tuned through the phase of the
interdot tunnelling matrix element by adjusting the flux
imbalance. Second, the conductance and the local densities of
states consist of the Breit-Wigner and the Fano resonances. The
corresponding Fano factors, the positions and the widths of these
resonances depend not only on the total magnetic flux, but also on
the magnetic flux imbalance. Thus electron transport can be
controlled by changing both the total magnetic flux and the
magnetic flux imbalance. Third, the normal AB oscillations with a
period of 2$\pi$ are destroyed, and complex periodic oscillations
can be generated. The oscillation periods for the total magnetic
flux and the magnetic flux imbalance are in general $4\pi$, while
in some particular situations, the $2\pi$-periodicity can be
recovered. Besides the $2\pi$ and the $4\pi$-period oscillation,
if $\phi$ and $\delta\phi$ are not independent, the oscillating
periods can have other possibilities. For the particular case
where the ratio of the magnetic flux in two subrings is an integer
$n$,\cite{Jiang02} the oscillating period becomes $2(n+1)\pi$. All
of these results can be easily read off from our general analytic
expressions. Furthermore, the AB oscillations can be very
sensitive to the magnetic flux imbalance. Thus accurate control of
the distribution of the magnetic flux is necessary for any
practical application of such an AB interferometer.

The paper is organized as follows. We describe the model in
Sec.~II. The general expressions of the differential conductance
and the density of states are derived there. In Sec.~III, we
evaluate the conductance as a function of the Fermi energy, and
show that the conductance in the present case consists of two
resonances which are composed of a Breit-Wigner resonance and a
Fano resonance. The local densities of states are calculated in
Sec.~IV, which show similar behaviors to the conductance. In
Sec.~V, the AB oscillation of the conductance as a function of
total magnetic flux and flux imbalance are studied. Finally, the
results are summarized and discussed in Sec.~VI. In Sec.~III-V
symmetric coupling of the dots to the left and right leads is
assumed for simplicity. The effect of asymmetric coupling is
briefly discussed in Appendix.

\begin{figure}
\includegraphics[width=3.in]{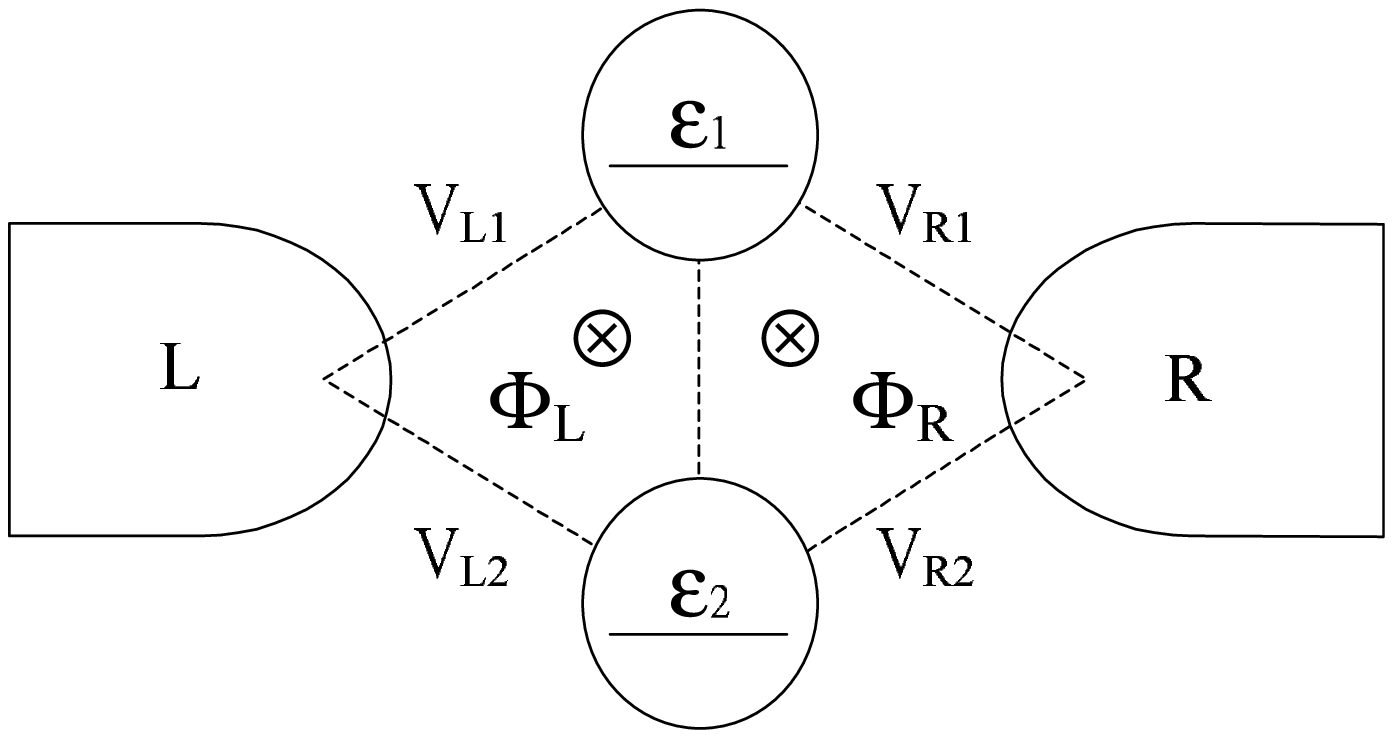}
\caption{Schematic diagram of two transversely coupled quantum
dots embedded in an Aharonov-Bohm interferometer.} \label{fig1}
\end{figure}

\section{Theory}

We consider an AB geometry as depicted in Fig.~\ref{fig1}, which
is basically equivalent to the experimental setup of
Ref.~\onlinecite{holleitner01}. The interdot and the intradot
electron-electron interactions are neglected, and only one energy
level in each dot is assumed relevant. The magnetic flux threading
the right-handed (left-handed) subring is denoted by $\Phi_R$
($\Phi_L$). Thus the total magnetic flux through the whole AB
interferometer is $\Phi=\Phi_L+\Phi_R$. The Hamiltonian of the
system can be written as
\begin{eqnarray}
H&=&\sum_{k\alpha}\varepsilon_{k\alpha}c_{k\alpha}^\dagger
c_{k\alpha} + {\bf d}^\dagger {\bf H}_0 {\bf d}\nonumber\\
&&+\sum_{k\alpha} \left( c_{k\alpha}^\dagger {\bf V}_{k\alpha}
{\bf d}+ {\rm H.c.} \right), \label{Hamiltonian}
\end{eqnarray}
where $c_{k\alpha}^\dagger$ ($c_{k\alpha}$) are the creation
(annihilation) operators for electrons with momentum $k$ in the
leads $\alpha=L, R$. For convenience, we introduce the following
matrix representation for the dynamics of the isolate double QD's
and the tunnelling between dots and leads:
\begin{eqnarray}
&&{\bf d}^\dagger=(d_{1}^\dagger, d_{2}^\dagger), \qquad {\bf
d}=\left(
\begin{array}{c}
d_{1}\\ d_{2}
\end{array}\right),
\nonumber\\
&&{\bf H}_0=\left(
\begin{array}{cc}
\varepsilon_1 & t \\ t^* & \varepsilon_2
\end{array}\right), \nonumber\\
&&{\bf V}_{k\alpha}=(V_{k\alpha 1} , V_{k\alpha 2} ), \nonumber
\end{eqnarray}
where $d_{i}$ ($d_{i}^\dagger$) annihilates (creates) an electron
in $i$-th dot ($i=1, 2$). The energy level in dot $i$ is denoted
by $\varepsilon_i$, which can be varied by the applied gate
voltages. $t$ is the interdot tunnelling coupling, and $V_{k\alpha
i}$ are the tunnelling matrix elements between dots and leads. The
magnetic flux is described by an AB phase factors attached to the
interdot tunnelling coupling and the tunnelling matrix elements.
We choose a gauge such that $t=|t| \exp(i\delta\phi/2)$, $V_{k L
1} = |V_{k L 1}| \exp(-i\phi/4)$, $V_{k L 2} = |V_{k L 2}|
\exp(i\phi/4)$, $V_{k R 2} = |V_{k R 2}| \exp(-i\phi/4)$, and
$V_{k R 1} = |V_{k R 1}| \exp(i\phi/4)$ with the (dimensionless)
total magnetic flux $\phi \equiv 2\pi \Phi/\Phi_0$ and the
(dimensionless) magnetic flux imbalance $\delta\phi \equiv 2\pi
(\Phi_L -\Phi_R)/\Phi_0$, where $\Phi_0 = h/e$ is the flux
quantum.

By employing the Landauer formula at zero temperature, the
differential conductance $G$ is related to the transmission
$T(\omega)$ of an electron of energy $\omega$,\cite{Meir}
\begin{equation}\label{conductance1}
G=\frac{e^{2}}{h} T(\varepsilon_F),
\end{equation}
where $\varepsilon_F$ stands for the Fermi level of both leads.
The total transmission $T(\omega)$ can be expressed as
\begin{equation}\label{transmission}
T(\omega)={\rm Tr} \left\{ {\bf \Gamma}^{L}{\bf G}^a(\omega){\bf
\Gamma}^{R}{\bf G}^r(\omega) \right\},
\end{equation}
where ${\bf G}^{r(a)}(\omega)$ is the Fourier transform of the
retarded (advanced) Green's function of the QD's, ${\bf
G}^{r(a)}(t)=\mp i\theta(\pm t)\langle\{{\bf d}(t),{\bf
d}^\dagger(0)\}\rangle$, where $\theta (t)$ is the step function
and the upper (lower) signs correspond to the retarded (advanced)
one. The matrix ${\bf\Gamma}^{\alpha}=2\pi\sum_k {\bf
V}_{k\alpha}^\dagger{\bf V}_{k\alpha}
\delta(\omega-\varepsilon_{k\alpha})$ describes the tunnelling
coupling of the two QD's to the lead $\alpha$. Here we neglect the
energy dependence of ${\bf \Gamma}^{\alpha}$. Notice that the
off-diagonal matrix elements of ${\bf\Gamma}^{\alpha}$ are complex
numbers due to the AB phase factors.

By using the equation of motion method, the exact retarded
(advanced) Green's function of the QD's is given by
\begin{widetext}
\begin{eqnarray}
{\bf G}^{r(a)}(\omega)=\frac{1}{D^{r(a)}(\omega)}\left(
\begin{array}{cc}
\omega-\varepsilon_2\pm \frac{i}{2}\Gamma_{22} & t \mp \frac{i}{2}
\Gamma_{12}
\\ t^* \mp \frac{i}{2} \Gamma_{21} & \omega-\varepsilon_1\pm \frac{i}{2}
\Gamma_{11}
\end{array}
\right) , \label{Green1}
\end{eqnarray}
where
\begin{eqnarray}
D^{r(a)}(\omega)&=&(\omega-\varepsilon_1\pm\frac{i}{2}{\Gamma}_{11})
(\omega-\varepsilon_2\pm\frac{i}{2}\Gamma_{22})
+\frac{1}{4}(|\Gamma^{L}_{12}|^2 +|\Gamma^{R}_{12}|^2) \nonumber\\
&&+\frac{1}{2}|\Gamma^{L}_{12}||\Gamma^{R}_{12}|\cos\phi-|t|^2 \mp
i|t||\Gamma^{L}_{12}|\cos(\frac{\phi + \delta\phi}{2}) \mp
i|t||\Gamma^{R}_{12}| \cos(\frac{\phi - \delta\phi}{2})
\label{Green2}
\end{eqnarray}
\end{widetext}
with ${\bf \Gamma}=\sum_\alpha{\bf \Gamma}^{\alpha}$. From the
above expression, we find that the conductance depends not only on
the total magnetic flux $\phi$, but also on the magnetic flux
imbalance $\delta\phi$ between the right and the left parts of the
present double-dot AB interferometer. As mentioned before, without
the interdot tunnelling coupling $|t|=0$, there is only one loop
in the AB interferometer, and the transport is determined only by
the phase $\phi$. In the special case of zero magnetic field, our
result reduces to that obtained in Ref.~\onlinecite{Guevara}.

From Eqs.~(\ref{Green1}) and (\ref{Green2}), general expressions
of the conductance and the local density of states can be reached.
However, since we focus our attention on the effect of flux
imbalance, we assume for simplicity that the magnitudes of
tunnelling matrix elements between the dots and the leads are the
same. Thus all of the magnitudes $|\Gamma^{\alpha}_{ij}|$ become
identical, which is denoted by $\Gamma/2$, while their values are
not the same because of the AB phase factors. (As shown in
Appendix, the following results are qualitatively unchanged even
when the magnitudes of $|\Gamma^{\alpha}_{ij}|$ are different.)
After substituting Eqs.~(\ref{transmission})-(\ref{Green2}) into
Eq.~(\ref{conductance1}), we can obtain a compact form of the
differential conductance
\begin{equation}\label{conductance2}
G= \frac{e^{2}}{h} \frac{(e_+ - e_-)^2 + 4\Delta}
{|(i-e_+)(i-e_-)-\Delta^\prime|^2}
\end{equation}
with
\begin{eqnarray}
e_{\pm} &\equiv& \frac{2\left(\bar{\varepsilon}\pm |t|
\cos\frac{\delta\phi}{2} -\varepsilon_F\right)}{\Gamma_{\pm}}, \nonumber\\
\Delta &\equiv& \frac{(\delta\varepsilon)^2}
{\Gamma_+\Gamma_-},  \nonumber\\
\Delta^\prime &\equiv& \frac{(\delta\varepsilon)^2+ 4|t|^2
\sin^2\frac{\delta\phi}{2}}{\Gamma_+\Gamma_-}, \label{parameter}\\
\Gamma_\pm &\equiv& (1\mp\cos\frac{\phi}{2})\Gamma,  \nonumber
\end{eqnarray}
where $\bar{\varepsilon}\equiv(\varepsilon_1+\varepsilon_2)/2$ and
$\delta\varepsilon\equiv\varepsilon_1-\varepsilon_2$ denote the
mean energy and the energy detuning of two QD's, respectively. The
parameters of $e_+$ and $\Gamma_+$ ($e_-$ and $\Gamma_-$) are
relevant to the antibonding (bonding) state of the QD molecule.
From the above result, we find that the conductance shows
oscillation patterns when either the total magnetic flux $\phi$ or
the magnetic flux imbalance $\delta\phi$ is changed (see also
Sec.~\ref{AB}). We note that Eq.~(\ref{conductance2}) becomes
identical to that obtained in Ref.~\onlinecite{Kang02} in the
special case of $\delta\phi=0$, which will even reduces to the
result obtained in Ref.~\onlinecite{Kubala02} in the case of the
absence of the interdot coupling ($t=0$). However, when the
distribution of the magnetic flux is non-uniform (i.e.,
$\delta\phi\neq 0$), more interesting behaviors can show up. It is
clear from the expression of $e_\pm$ that, by adjusting the phase
of the interdot tunnelling matrix element through the flux
imbalance $\delta\phi$ in the AB interferometer, one can tune the
overlapping of the dot's wave functions. Moreover, the level
crossing of the bonding and the antibonding states can occur by
varying $\delta\phi$. We emphasize again that these are possible
only when interdot tunnelling coupling is nonzero.

The local density of states at the $i$-th QD is given by
$\rho_i(\omega)=-{\rm Im}G_{ii}^r(\omega)/\pi$. By using the
expression of the retarded Green's function in Eq.~(\ref{Green1}),
the general formula of the local densities of states at
$\varepsilon_F$ can be written as
\begin{widetext}
\begin{eqnarray}\label{DOS}
\rho_{1(2)}(\varepsilon_F)=
\frac{\Gamma_+e_+^2+\Gamma_-e_-^2+(\Gamma_++\Gamma_-)(1+\Delta^\prime)\mp
2 \; \delta\varepsilon (e_+ + e_-)}
{2\pi\Gamma_+\Gamma_-|(i-e_+)(i-e_-)-\Delta^\prime|^2},
\end{eqnarray}
\end{widetext}
where the upper (lower) sign corresponds to $\rho_1$ ($\rho_2$).
Because most of the parameters defined in Eq.~(\ref{parameter})
depend on $\phi$ and/or $\delta\phi$, the local densities of
states $\rho_1$ and $\rho_2$ again have oscillating behaviors as
the total magnetic flux and/or the magnetic flux imbalance are
varied.

\section{conductance}

The general form of the conductance in Eq.~(\ref{conductance2}) is
quite similar to that obtained in Ref.~\onlinecite{Kang02} for the
$\delta\phi=0$ case. Therefore, following the same kind of
analysis, one can easily show that the conductance in the present
case consists of two resonances which are composed of a
Breit-Wigner resonance and a Fano resonance.

Without loss of generality, we can discuss the case in the limit
$\Gamma_- \gg \Gamma_+$. If the energy scale is larger than
$\Gamma_+$ ($|e_+|\gg 1$), the conductance in
Eq.~(\ref{conductance2}) indeed takes the Breit-Wigner form:
\begin{equation}\label{G:BW}
G\simeq G_{\rm BW}=\frac{e^{2}}{h} \frac{1}{e_-^2 + 1}.
\end{equation}
Its width is $\Gamma_-$ which depends on the total magnetic flux
but not on the flux imbalance [see Eq.~(\ref{parameter})]. Near
the narrower resonance regime $(|e_+|\lesssim 1)$, the conductance
does show the Fano-resonance behavior,
\begin{equation}\label{G:Fano}
G\simeq G_{\rm Fano}=\frac{e^{2}}{h} T_b\frac{|e^\prime_+
+Q|^2}{{e^\prime_+}^2+1},
\end{equation}
where the background transmission is given by $T_b=1/(q^2+1)$ with
$q= 4|t|\cos(\delta\phi/2)/\Gamma_-$, and
\begin{equation}
e^\prime_+ = \frac{e_+ + q T_b \Delta^\prime}{1 + T_b
\Delta^\prime}.
\end{equation}
The modified Fano factor is now given by
\begin{equation}
Q=q\frac{1-T_b\Delta^\prime}{1+T_b\Delta^\prime} +
i\frac{2\sqrt{\Delta}}{1+T_b\Delta^\prime}, \label{eq:Fano1}
\end{equation}
and the width of the Fano resonance becomes $\Gamma^\prime_+ =
\Gamma_+ (1+T_b\Delta^\prime)$. While these results are formally
identical to those obtained in Ref.~\onlinecite{Kang02}, we show
that the modified Fano factor, the position and the width of the
resonance all depend not only on the total magnetic flux $\phi$,
but also on the magnetic flux imbalance $\delta\phi$. Thus
transport signals can be manipulated by adjusting both $\phi$ and
$\delta\phi$.

For the perfectly symmetrical geometry (i.e, $\delta\varepsilon=0$
and $|\Gamma^{\alpha}_{ij}|$ are all the same) and in the case of
zero magnetic field (or more generally $|\cos(\phi/2)|=1$ and
$|\cos(\delta\phi/2)|=1$), the width of the Fano resonance becomes
zero (or the life time of the antibonding state becomes infinitely
long). It is because the antibonding state now becomes totally
decoupled to the leads. Therefore, the Fano resonance will
disappear in this case. This phenomena had been pointed out a
decade ago,\cite{Shahbazyan94} which is recently called as a
``ghost of Fano resonance".\cite{Guevara} We note that this
disappearance of the Fano resonance can happen only in this very
special case. For example, even for the perfectly symmetrical
geometry, the Fano resonance will show up when the magnetic field
is turned on (see also Appendix).

\begin{figure}
\includegraphics[width=3.in]{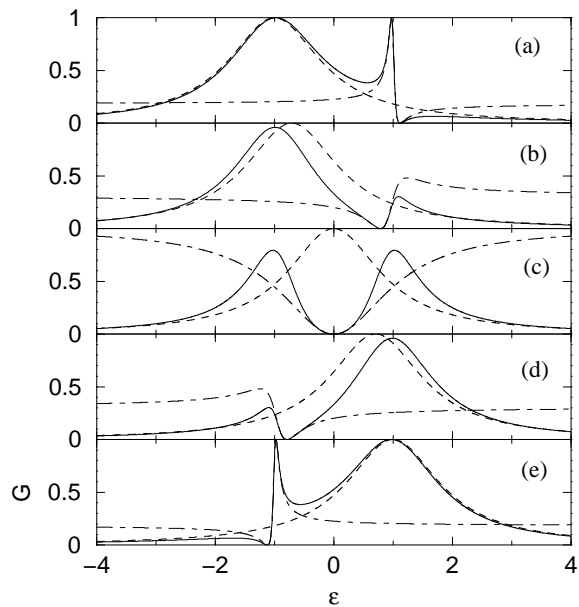}
%
\caption{Differential conductance $G$ in unit of $e^2 /h$ (solid
lines) as a function of the Fermi energy (in unit of $\Gamma$) for
various flux imbalance. The results shown in panels (a) through
(e) correspond to the flux imbalance $\delta\phi=0$, $\pi/2$,
$\pi$, $3\pi/2$, and $2\pi$. Other parameters are given by
$\bar{\varepsilon}=\delta\varepsilon=0$, $|t|/\Gamma=1$, and
$\phi=0.3\pi$. Dashed and dashed-dotted lines denote the
Breit-Wigner and the generalized Fano asymptote given in
Eq.(\ref{G:BW}) and Eq.(\ref{G:Fano}), respectively.}
\label{fig:fano}
\end{figure}

To illustrate the above discussions, the differential conductance
$G$ as a function of the Fermi energy $\varepsilon_F$ is shown in
Fig.~\ref{fig:fano} for various $\delta\phi$ with $|t|/\Gamma=1$,
$\delta\varepsilon=0$ (the so-called ``covalent limit"), and
$\phi=0.3\pi$. Here $\bar{\varepsilon}$ is taken as the
zero-energy level. Fig.~\ref{fig:fano}(a) reproduces the topmost
one of Fig.~2 in Ref.~\onlinecite{Kang02}. We find that, as
$\delta\phi$ increases from zero to $\pi$, two resonances come
closer and closer, and finally two energy level of resonance meet
each other when $\delta\phi=\pi$. This can be understood from the
expressions of $e_-$ and $e^\prime_+$. Further increase of
$\delta\phi$, the Breit-Wigner resonance keeps moving to the
positive-energy side, while the Fano resonance goes to the
negative-energy side. The resonance levels will move back when
$\delta\phi >2\pi$ and the curve for $\delta\phi=0$ is recovered
when $\delta\phi$ is increased to $4\pi$. Thus the conductance has
in general a period of $4\pi$ for $\delta\phi$ (for further
discussions, see Sec.~\ref{AB}).

In Fig.~\ref{fig:fano}, we find that the zero and the full
transmission can occur at some particular values of the Fermi
energies. The analytic expressions of these Fermi energies can be
easily derived. From Eqs.~(\ref{conductance2}) and
(\ref{parameter}), it is obvious that, if the two levels of the
double QD's are not the same ($\delta\varepsilon\neq 0$), the
conductance cannot be zero. In this case, $\Delta\neq 0$ and the
modified Fano factor in Eq.~(\ref{eq:Fano1}) becomes a complex
number. It means that the completely destructive interference in
the present AB interferometer will not appear in this case.
However, when $\delta\varepsilon=0$, the completely destructive
interference and transmission zero happen at the Fermi energy
\begin{equation}\label{T0}
\varepsilon_F=\bar{\varepsilon} +
|t|\cos\frac{\delta\phi}{2}/\cos\frac{\phi}{2},
\end{equation}
provided that $\sin(\phi/2)\neq 0$ or $\sin(\delta\phi/2)\neq 0$.
It is sensitive to the total magnetic flux $\phi$ and the magnetic
flux imbalance $\delta\phi$. On the other hand, the conductance
can reach its quantum limit $G=e^2/h$ at the Fermi energies
\begin{equation}\label{T1}
\varepsilon_F=\bar{\varepsilon} \pm
\sqrt{\left(\frac{\delta\varepsilon}{2}\right)^2 +|t|^2
-\left(\frac{\Gamma}{2}\right)^2 \sin^2\frac{\phi}{2}}
\end{equation}
if $\sin(\phi/2)\sin(\delta\phi/2)=0$ and the expression in the
square root is positive. The last term in the square root of
Eq.~(\ref{T1}) gives the so-called ``flux-dependent level
attraction" mentioned in Ref.~\onlinecite{Kubala02}. From the
above discussion, it is realized that the value of transmission
will in general not be zero or one unless some particular
conditions happen to be satisfied.\cite{note1}

\section{local density of states}

Similar behaviors to the results in the previous section can be
found by examining the local density of states in each of the
quantum dots.

For the perfectly symmetrical geometry, $\delta\varepsilon=0$ and
$\delta\phi=0$, therefore $\Delta=\Delta^\prime=0$,
Eq.~(\ref{DOS}) reduces to
\begin{equation}
\rho_{1}=\rho_{2}
=\frac{1}{2\pi\Gamma_-(e^2_-+1)}+\frac{1}{2\pi\Gamma_+(e^2_++1)}.
\end{equation}
That is, both of the local densities of states take the form of
the superposition of two Breit-Wigner resonances of widths
$\Gamma_-$ and $\Gamma_+$ at the bonding and antibonding energies,
respectively. However, in other cases, following the same kind of
analysis in the previous section, it can be shown that the local
densities of states consist of a Breit-Wigner at the bonding
energy and a Fano line shape at the antibonding energy.

\begin{figure}
\includegraphics[width=3.in]{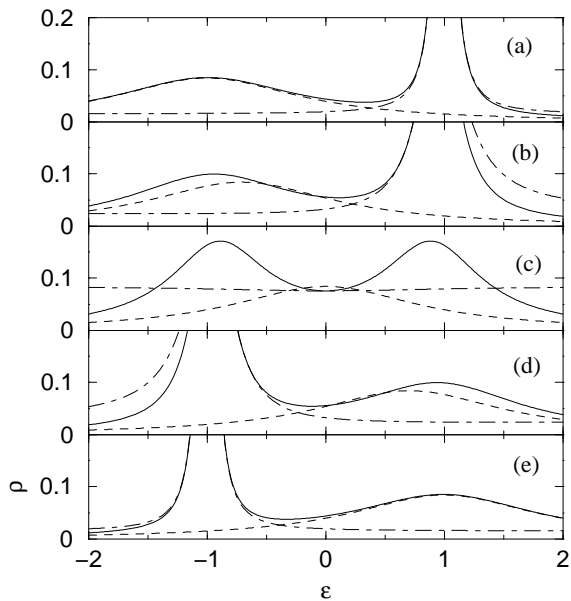}
%
\caption{Local density of states at a QD as a function of the
Fermi energy (in unit of $\Gamma$). The results shown in panels
(a) through (e) correspond to the flux imbalance $\delta\phi=0$,
$\pi/2$, $\pi$, $3\pi/2$, and $2\pi$. Other parameters are the
same as those used in Fig.~\ref{fig:fano}. Dashed and
dashed-dotted lines denote the Breit-Wigner and the generalized
Fano asymptote given in Eq.(\ref{DOS:BW}) and Eq.(\ref{DOS:Fano}),
respectively.} \label{fig:DOS}
\end{figure}

Without loss of generality, we discuss the case again in the limit
$\Gamma_- \gg \Gamma_+$. If the energy scale is larger than
$\Gamma_+$ ($|e_+|\gg 1$), both of the local densities of states
in Eq.~(\ref{DOS}) take the Breit-Wigner form of width $\Gamma_-$:
\begin{equation}\label{DOS:BW}
\rho_{1(2)}\simeq\frac{1}{2\pi \Gamma_-(e_-^2+1)}.
\end{equation}
Near the narrower resonance regime $(|e_+|\lesssim 1)$, the local
densities of states show the Fano-resonance behavior,
\begin{equation}\label{DOS:Fano}
\rho_{1(2)}\simeq \rho_b\frac{|e^\prime_+
-P_{1(2)}|^2}{1+{e^\prime_+}^2},
\end{equation}
where $\rho_b\equiv T_b/(2\pi\Gamma_-)$ and the corresponding Fano
factor is
\begin{widetext}
\begin{equation}\label{eq:Fano2}
P_{1(2)}=\frac{qT_b\Delta^\prime\pm
\sqrt{\Delta\frac{\Gamma_-}{\Gamma_+}}}{1+T_b\Delta^\prime}
+i\frac{\sqrt{\left[ \left(q\pm
\sqrt{\Delta\frac{\Gamma_+}{\Gamma_-}}\right)^2 +
(1+\Delta^\prime-\Delta) \left(1+\frac{\Gamma_+}{\Gamma_-}\right)
\right] \frac{\Gamma_-}{\Gamma_+}}}{1+T_b\Delta^\prime},
\end{equation}
\end{widetext}
where the upper (lower) sign corresponds to $P_1$ ($P_2$). The
Fano factor in Eq.~(\ref{eq:Fano2}) is more complicated than that
for the conductance [Eq.~(\ref{eq:Fano1})]. Thus it is possible in
some situations (say, $\Delta=0$) that $P_{1(2)}$ is a complex
number but $Q$ is real. Notice that the present Fano factors are
in general not the same for different QD's. From the above
results, it can be understood that the Fano factor $P_{1(2)}$, the
position and the width of the resonance all can be manipulated by
adjusting both of the total magnetic flux $\phi$ and the magnetic
flux imbalance $\delta\phi$. In the absence of the magnetic field,
the above expressions are equivalent to Eqs.~(27) and (28) of
Ref.~\onlinecite{Guevara}.

The local density of states of as a function of Fermi energy
$\varepsilon_F$ is plotted in Fig.~\ref{fig:DOS} for the same
parameters used in Fig.~\ref{fig:fano}. In this case,
$\rho_1=\rho_2$. We find that same level crossing appears as
$\delta\phi$ is varied and the curve has again a period of $4\pi$
for $\delta\phi$ (for further discussions, see Sec.~\ref{AB}). We
notice that the local density of states is always nonvanishing for
the chosen parameters, because the Fano factor $P_{1(2)}$ is
complex in these cases. As compared with Fig.~\ref{fig:fano}, it
is found that the states with small density of states can have
almost full transmission and those with large density of states
can show zero transmission. This indicates that the full and the
zero transmission are indeed consequences of the quantum
interference, as mentioned before.

\section{Aharonov-Bohm Oscillations}
\label{AB}

We now discuss the AB oscillation of the conductance as a function
of total magnetic flux $\phi$ and the magnetic flux imbalance
$\delta\phi$ with fixed mean energy $\bar{\varepsilon}$ and energy
detuning $\delta\varepsilon$ of two QD's.

From Eq.~(\ref{conductance2}), it is clear that the conductance
[and also the local densities of states, see Eq.~(\ref{DOS})] is a
periodic function of both $\phi$ and $\delta\phi$. The oscillation
periods for $\phi$ and $\delta\phi$ are in general $4\pi$. The
$4\pi$-period oscillation for $\delta\phi$ has be implied in
Fig.~\ref{fig:fano}, and the $4\pi$-period oscillation for $\phi$
in the case of $\delta\phi=0$ has be found in
Ref.~\onlinecite{Kang02}. However, in the following particular
situations, the $2\pi$-periodicity can occur. (i) If we take
$\varepsilon_F=\bar{\varepsilon}$, then $\Gamma_+ e_+ = -\Gamma_-
e_- = 2|t|\cos(\delta\phi/2)$. In this case, one can show that the
conductance is now a function of $\cos^2(\phi/2)$ and
$\cos^2(\delta\phi/2)$. Hence the conductance shows $2\pi$-period
oscillation.\cite{note2} This $2\pi$-period oscillation for
$\delta\phi$ can be understood from Fig.~\ref{fig:fano}, if we
trace the change of $G$ at $\varepsilon_F=\bar{\varepsilon}=0$ as
$\delta\phi$ is varied from 0 to $2\pi$. (ii) If $\delta\phi=\pi$
[or more generally $\cos(\delta\phi/2)=0$], we have $\Gamma_+ e_+
= \Gamma_- e_- = 2\left(\bar{\varepsilon}-\varepsilon_F\right)$.
Therefore, the conductance becomes a function of $\cos^2(\phi/2)$
and the oscillation periods for $\phi$ is $2\pi$. (iii) If
$\phi=\pi$ [or more generally $\cos(\phi/2)=0$], we have $\Gamma_+
= \Gamma_- = \Gamma$. In this case, the conductance is a function
of $\cos^2(\delta\phi/2)$ with a $2\pi$-period oscillation for
$\delta\phi$.

\begin{figure}
\includegraphics[width=3in]{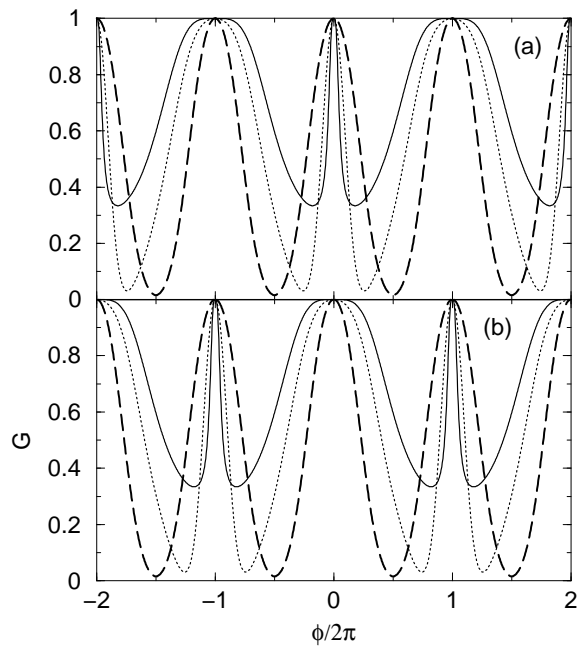}
%
\caption{AB oscillations as a function of the total magnetic flux
$\phi$ for various flux imbalance $\delta\phi$. (a) $\delta\phi=0$
(solid line), $\pi/2$ (dotted line), $\pi$ (dashed line). (b)
$\delta\phi=\pi$ (dashed line), $3\pi/2$ (dotted line), $2\pi$
(solid line). Here the Fermi energy is taken as the energy for the
bonding state when the QD's are decoupled to the leads (see text).
Other parameters are given by $\bar{\varepsilon}=0$,
$|t|/\Gamma=0.3$, $\delta\varepsilon/\Gamma=0.1$. } \label{fig:AB}
\end{figure}

As an illustration, the AB oscillation as a function of $\phi$ is
shown in Fig.~\ref{fig:AB} for different values of $\delta\phi$
with $|t|/\Gamma=0.3$ and $\delta\varepsilon/\Gamma=0.1$. Here we
choose $\varepsilon_F = \bar{\varepsilon}
-\sqrt{\left(\delta\varepsilon/2\right)^2 +|t|^2}$, which is the
energy for the bonding state when the QD's are decoupled to the
leads. $\bar{\varepsilon}$ is again taken as the zero-energy level
for convenience. The curve for $\delta\phi=0$ corresponds to the
$A1$ curve in Fig.~4(c) of Ref.~\onlinecite{Kang02}, where sharp
peaks around $\phi=4n\pi$ ($n=0, \pm 1, \pm 2, \dots$) result. It
shows that the conductance can be very sensitive to the total
magnetic flux $\phi$ for the chosen parameters (say, near
$\phi=0$). This opens the possibility to manipulate transport in a
nontrivial way by varying the magnetic field. As shown in
Fig.~\ref{fig:AB}, we see that this sensitivity can be even
strengthened when $\delta\phi$ is present. As $\delta\phi$
increases to $\pi$, the periodicity even changes from $4\pi$ to
$2\pi$, as discussed above.

Besides the $2\pi$ and the $4\pi$-period oscillation studied
above, it is found that, if $\phi$ and $\delta\phi$ are not
independent, the oscillating periods can have other
possibilities.\cite{Jiang02}  The authors of
Ref.~\onlinecite{Jiang02} show numerically that the oscillating
period will be $2(n+1)\pi$ when $\delta\phi=[(n-1)/(n+1)]\phi$ (or
the magnetic flux ratio $\Phi_L / \Phi_R =n$) with $n$ being
integers. This result can be easily explained from our analytic
expression of Eq.~(\ref{conductance2}). As $\phi$ being varied
from 0 to $2(n+1)\pi$, $\delta\phi$ is increased from 0 to
$2(n-1)\pi$, therefore we have $\Gamma_\pm \rightarrow
\Gamma_\pm$, $e_\pm \rightarrow e_\pm$ ($\Gamma_\pm \rightarrow
\Gamma_\mp$, $e_\pm \rightarrow e_\mp$), and $\Delta^\prime
\rightarrow \Delta^\prime$ if $n$ is odd (even). This makes $G$
back to its original value. It means that the oscillating period
is $2(n+1)\pi$ in this case.

\section{Conclusions}

In conclusion, we have investigated the influence of the
non-uniform distribution of the magnetic flux on quantum transport
through coupled double QD's embedded in an AB interferometer. We
show that the effective tunnelling coupling between two dots can
be tuned by the magnetic flux imbalance $\delta\phi$ threading two
AB subrings. Therefore, the conductance and the local densities of
states become periodic functions of $\delta\phi$. Moreover, the
conductance and the local densities of states are shown to be
composed of a Breit-Wigner resonance and a Fano resonance. The
corresponding Fano factors, the positions and the widths of the
resonances all depend not only on $\phi$, but also on
$\delta\phi$. Thus transport signals can be manipulated by
adjusting both $\phi$ and $\delta\phi$. Finally, we point out that
the AB oscillations can be very sensitive to $\delta\phi$. Thus
accurate control of the distribution of the magnetic flux is
necessary for any practical application of such an AB
interferometer.

\begin{acknowledgments}
Z.M.B. work is financially supported partly by Grant No.
NSC-91-2816-M-029-0002-6. M.F.Y. is supported by Grant No. NSC
91-2112-M-029-007. Y.C.C. acknowledges financial support by Grant
No. NSC 91-2112-M-029-006.
\end{acknowledgments}

\appendix*
\section{effect of asymmetric coupling between dots and leads}

In this appendix, we devote ourselves to the effect of difference
in the matrix elements of $\Gamma^{\alpha}_{ij}$. As an example,
we follow the setup which are considered in
Ref.~\onlinecite{Guevara} in the case of zero magnetic field:
$\Gamma_{11}^{R}=\Gamma_{22}^{L}\equiv \gamma_{1}$ and
$\Gamma_{11}^{L}=\Gamma_{22}^{R}\equiv \gamma_{2}$, and the
magnitudes of the off-diagonal matrix elements
$|\Gamma_{21}^{L}|=|\Gamma_{12}^{L}|=|\Gamma_{21}^{R}|=|\Gamma_{12}^{R}|\equiv
\sqrt{\gamma_{1}\gamma_{2}}$.

In this case, the conductance and the local densities of states
are again given by Eq.~(\ref{conductance2}) and (\ref{DOS})
respectively, where $e_{\pm}$ and $\Delta^\prime$ are identical to
those given in Eq.~(\ref{parameter}), but $\Delta$ and
$\Gamma_{\pm}$ now become:
\begin{eqnarray}
\Delta&=&\left[ \frac{
2(\gamma_1-\gamma_2)|t|\sin\frac{\delta\phi}{2}
-2\sqrt{\gamma_1\gamma_2} \; \delta\varepsilon\sin\frac{\phi}{2}
}{\Gamma_+\Gamma_-} \right]^2 ,  \\
\Gamma_{\pm}&=&(\gamma_1+\gamma_2)\mp
2\sqrt{\gamma_1\gamma_2}\cos\frac{\phi}{2}  . \label{Gamma}
\end{eqnarray}
When $\gamma_1=\gamma_2\equiv \Gamma/2$, the above expressions
reduce to the corresponding ones in Eq.~(\ref{parameter}). Thus,
merely by replacing the functional forms of $\Delta$ and
$\Gamma_{\pm}$, the discussions in the text are still applied in
this case of asymmetric coupling between dots and leads. From
Eq.~(\ref{Gamma}), one finds that, when $\phi=0$, $\Gamma_{+}$
approaches zero as $\gamma_1 - \gamma_2 \rightarrow 0$. This
result had been pointed out in Ref.~\onlinecite{Shahbazyan94}.
However, for nonzero magnetic field, none of $\Gamma_{\pm}$ will
be zero in the limit $\gamma_1 - \gamma_2 \rightarrow 0$ as long
as $|\cos\frac{\phi}{2}| \neq 1$. Thus the phenomena of a ``ghost
of Fano resonance"\cite{Guevara} will not appear when the magnetic
field is nonzero.


\begin{thebibliography}{150}

\bibitem{books}
{\em Mesoscopic~Electron~Transport}, edited by L. L.~Sohn, L.
P.~Kouwenhoven, and G.~Sch\"on (Kluwer, Dordrecht,1997); {\em
Single Charge Tunnelling}, edited by H.~Grabert and M. H.~Devoret,
(Plenum, New York, 1991); D. V.~Averin and K. K.~Likharev, in {\em
Mesoscopic Phenomena in Solids}, ed. B. L.~Altshuler, P. A.~Lee,
and R. A.~Webb (Elsevier, Amsterdam, 1991), pp. 173-271. 

\bibitem{Yacobi95}
A. Yacoby, M. Heiblum, D. Mahalu, and H. Shtrikman, Phys. Rev.
Lett. {\bf 74}, 4047 (1995).

\bibitem{Schuster97}
R. Schuster, E. Buks, M. Heiblum, D. Mahalu, V. Umansky, and H.
Shtrikman, Nature (London) {\bf 385}, 417 (1997).

\bibitem{Ji00}
Y. Ji, M. Heiblum, D. Sprinzak, D. Mahalu, and H. Shtrikman,
Science {\bf 290}, 779 (2000); Y. Ji, M. Heiblum, and H.
Shtrikman, Phys. Rev. Lett. {\bf 88}, 076601 (2002).

\bibitem{Wiel00}
W.G. van der Wiel, S. De Franceschi, T. Fujisawa, J.M. Elzerman,
S. Tarucha, and L.P. Kouwenhoven, Science {\bf 289}, 2105 (2000).

\bibitem{Kobayashi}
K. Kobayashi, H. Aikawa, S. Katsumoto, and Y. Iye, J. Phys. Soc.
Jpn. {\bf 71}, L2094-2097 (2002); H. Aikawa, K. Kobayashi, A.
Sano, S. Katsumoto, and Y. Iye, cond-mat/0309084.

\bibitem{Kobayashi-fano}
K. Kobayashi, H. Aikawa, S. Katsumoto, and Y. Iye, Phys. Rev.
Lett. {\bf 88}, 256806 (2002); cond-mat/0309570.

\bibitem{holleitner01}
A. W. Holleitner, C. R. Decker, H. Qin, K. Eberl, and R. H. Blick,
Phys. Rev. Lett. {\bf 87}, 256802 (2001).

\bibitem{holleitner02}
A. W. Holleitner, R. H. Blick, A. K. H\"uttel, K. Eberl, and J. P.
Kotthaus, Science {\bf 297}, 70 (2002).

\bibitem{blick03}
R. H. Blick, A. K. H\"{u}ttel, A. W. Holleitner, E. M.
H\"{o}hberger, H. Qin, J. Kirschbaum, J. Weber, W. Wegscheider, M.
Bichler, K. Eberl, and J. P. Hotthaus, Physica E {\bf 16}, 76
(2003).

\bibitem{Sigrist03}
M. Sigrist, A. Fuhrer, T. Ihn, K. Ensslin, W. Wegscheider, M.
Bichler, cond-mat/0307269; M. Sigrist, A. Fuhrer, T. Ihn, K.
Ensslin, S. E. Ulloa, W. Wegscheider, M. Bichler,
cond-mat/0308223.

\bibitem{Loss00}
D. Loss and E.V. Sukhorukov, Phys. Rev. Lett. {\bf 84}, 1035
(2000).

\bibitem{Jiang02}
Z. T. Jiang, J. Q. You, S. B. Bian, and H. Z. Zheng, Phys. Rev. B
{\bf 66}, 205306 (2002)

\bibitem{Kang02}
Kicheon Kang and Sam Young Cho, cond-mat/0210009.

\bibitem{note-1}
Here we suggest a possible way to adjust experimentally the flux
imbalance in the double-dot AB interferometer. Periodic magnetic
field with period about one micrometer has been generated by using
a regular array of superconductor [H. A. Carmona {\it et al.},
Phys. Rev. Lett. {\bf 74}, 3009 (1995)] or micromagnet [P.~D.~Ye
{\it et al.}, Phys. Rev. Lett. {\bf 74}, 3013 (1995)]. By covering
various periodic magnetic fields with different periods on the top
of the sample with the double-dot AB interferometer, one can
change to some extent the magnetic flux imbalance threading two
subrings.

\bibitem{note-2}
For example, our general analytic expressions of the conductance
and the local densities of states provide a better fitting formula
for the experimental data as long as the flux imbalance is there.
The fitting parameter of the flux imbalance can give an indication
of the degree of asymmetry in the construction of the AB
interferometer.

\bibitem{Guevara}
M. L. Ladr\'on de Guevara, F. Claro, and Pedro A. Orellana, Phys.
Rev. B {\bf 67}, 195335 (2003). There are some typographic errors
in their Eqs.~(9), (10), (14) and (15).

\bibitem{Meir}
Y. Meir and N. Wingreen, Phys. Rev. Lett. {\bf 68}, 2512 (1992);
A. P. Jauho, N. S. Wingreen, and Y. Meir, Phys. Rev. B {\bf 50},
5528 (1994).

\bibitem{Kubala02}
B. Kubala and J. K\"onig, Phys. Rev. B {\bf 65}, 245301 (2002).

\bibitem{Shahbazyan94}
T. V. Shahbazyan and M. E. Raikh, Phys. Rev. B {\bf 49}, 17~123
(1994).

\bibitem{note1}
Similar expressions of the Fermi energies corresponding to the
zero and the full transmission have be given in
Refs.~\onlinecite{Guevara} and \onlinecite{Kubala02} for various
limiting cases. By taking $\delta\phi=0$ and $t=0$, Eq.~(\ref{T1})
reduces to the condition of full transimission given in
Ref.~\onlinecite{Kubala02}. On the other hand, when the magnetic
field is absent and the setup is perfectly symmetrical, i.e.,
$\phi=\delta\phi=0$ and $\delta\varepsilon=0$, Eqs.~(\ref{T0}) and
(\ref{T1}) can be related to those in Ref.~\onlinecite{Guevara}.

\bibitem{note2}
In this case, the total density of states, $\rho_1 + \rho_2$, has
also $2\pi$ oscillating periods for $\phi$ and $\delta\phi$.
However, the oscillating periods for $\phi$ and $\delta\phi$ of
the {\it local} densities of states, $\rho_1$ and $\rho_2$, are
still $4\pi$ unless $\Delta=0$ (i.e., the enrrgy detuning
$\delta\varepsilon=0$).


\end{thebibliography}
\end{document}